\documentclass[twocolumn,aps,prl,superscriptaddress]{revtex4}

\usepackage{graphicx}
\usepackage{hyperref}

\begin{document}

\title{Generating Single Microwave Photons in a Circuit}


\author{A.~A.~Houck\footnote{Authors with contributed equally to this work.}}
\affiliation{Departments of Applied Physics and Physics, Yale
University, New Haven, CT 06520}
\author{D.~I.~Schuster$^{*}$}
\affiliation{Departments of Applied Physics and Physics, Yale
University, New Haven, CT 06520}
\author{J.~M.~Gambetta}
\affiliation{Departments of Applied Physics and Physics, Yale
University, New Haven, CT 06520}
\author{J.~A.~Schreier}
\affiliation{Departments of Applied Physics and Physics, Yale
University, New Haven, CT 06520}
\author{B.~R.~Johnson}
\affiliation{Departments of Applied Physics and Physics, Yale
University, New Haven, CT 06520}
\author{J.~M.~Chow}
\affiliation{Departments of Applied Physics and Physics, Yale
University, New Haven, CT 06520}
\author{J.~Majer}
\affiliation{Departments of Applied Physics and Physics, Yale
University, New Haven, CT 06520}
\author{L.~Frunzio}
\affiliation{Departments of Applied Physics and Physics, Yale
University, New Haven, CT 06520}
\author{M.~H.~Devoret}
\affiliation{Departments of Applied Physics and Physics, Yale
University, New Haven, CT 06520}
\author{S.~M.~Girvin}
\affiliation{Departments of Applied Physics and Physics, Yale
University, New Haven, CT 06520}
\author{R.~J.~Schoelkopf}
\affiliation{Departments of Applied Physics and Physics, Yale
University, New Haven, CT 06520}
\date{\today}

\begin{abstract}

Electromagnetic signals in circuits consist of discrete
photons\cite{schuster07}, though conventional voltage sources can
only generate classical fields with a coherent superposition of many
different photon numbers.  While these classical signals can control
and measure bits in a quantum computer (qubits), single photons can
carry quantum information, enabling non-local quantum interactions,
an important resource for scalable quantum
computing\cite{divincenzo00}.  Here, we demonstrate an on-chip
single photon source in a circuit quantum electrodynamics (QED)
architecture\cite{blais04}, with a microwave transmission line
cavity that collects the spontaneous emission of a single
superconducting qubit with high efficiency. The photon source is
triggered by a qubit rotation, as a photon is generated only when
the qubit is excited. Tomography of both qubit and fluorescence
photon shows that arbitrary qubit states can be mapped onto the
photon state, demonstrating an ability to convert a stationary qubit
into a flying qubit.  Both the average power and voltage of the
photon source are characterized to verify performance of the system.
This single photon source is an important addition to a rapidly
growing toolbox for quantum optics on a chip.
\end{abstract}

\maketitle

Numerous approaches to generating single photons, particularly
optical photons, have been proposed and demonstrated in recent
years\cite{oxborrow05}.  The underlying principle for generating
single photons from atoms or qubits is straightforward:  an excited
qubit can relax to its ground state by emitting a
photon\cite{clauser74, kimble77}, A pulse that excites the qubit can
therefore trigger a single photon emission, as long as the control
and emission photons can be separated. Early experiments
demonstrated this photon generation from single
ions\cite{diedrich87}, atoms\cite{darquie05},
molecules\cite{basche92,brunel99, lounis00}, nitrogen
vacancies\cite{kurtsiefer00}, and quantum
dots\cite{michler00_nature, pelton02}, though radiation in all
directions made efficient collection difficult.  In a cavity QED
source, the atom or qubit is coupled to a single photonic mode of a
cavity, enhancing the rate of decay to that mode through the Purcell
effect\cite{purcell46} and allowing a source where photons are
emitted into a controlled channel. Atoms \cite{brattke01, kuhn02,
mckeever04,maitre97}, ions \cite{keller04}, and quantum
dots\cite{moreau01, santori02,pelton02} have been used to generate
optical photons efficiently in this manner.

Here, we implement a cavity QED system in a circuit\cite{wallraff04,
chiorescu04}, where a superconducting qubit and transmission line
cavity are coupled such that the dominant channel for relaxation of
the qubit is to spontaneously emit a photon into the cavity. Each
time the qubit is excited, the most likely outcome is the generation
of one (and only one) photon at a random time, with the distribution
of times characterized by the decay rate of the qubit.  The
challenge is to create a system where spontaneous emission dominates
other relaxation channels. This spontaneous emission rate can be
determined from the Hamiltonian of the system, the well-known
Jaynes-Cummings Hamiltonian, $H = \hbar\omega_{\rm{a}}
\sigma_{\rm{z}}/2+\hbar\omega_{\rm{r}} (a^{\dagger}a+1/2)+\hbar g
(a^{\dagger} \sigma^{-}+a\sigma^{+}) $. The first two terms
represent a qubit with frequency $\omega_{\rm{a}}$ described by
Pauli operators $\sigma_{\rm{x}}$, $\sigma_{\rm{y}}$, and
$\sigma_{\rm{z}}$ and raising and lowering operators $\sigma^+$ and
$\sigma^-$, and a single photon mode of frequency $\omega_{\rm{r}}$
described by the photon operators $a$ and $a^{\dagger}$. The final
term represents a coupling of strength $g$ between the qubit and the
photon, which mixes the individual qubit and photon eigenstates.
When far detuned ($\Delta = \omega_{\rm{r}}-\omega_{\rm{a}} \gg g$),
the qubit acquires a small photonic component of the wavefunction,
of magnitude $g/\Delta$. This opens a new source of decay for the
qubit, as the photonic component of the qubit can decay at the
cavity decay rate, $\kappa$, resulting in a new qubit decay rate
$\gamma_{\kappa} = (g/\Delta)^2\kappa$. The qubit can be an
efficient photon source if this new decay rate dominates over other
non-radiative decay rates, $\gamma_{\kappa}
> \gamma_{\rm{NR}}$.

\begin{figure}[!bp]

\includegraphics[width=0.48\textwidth]{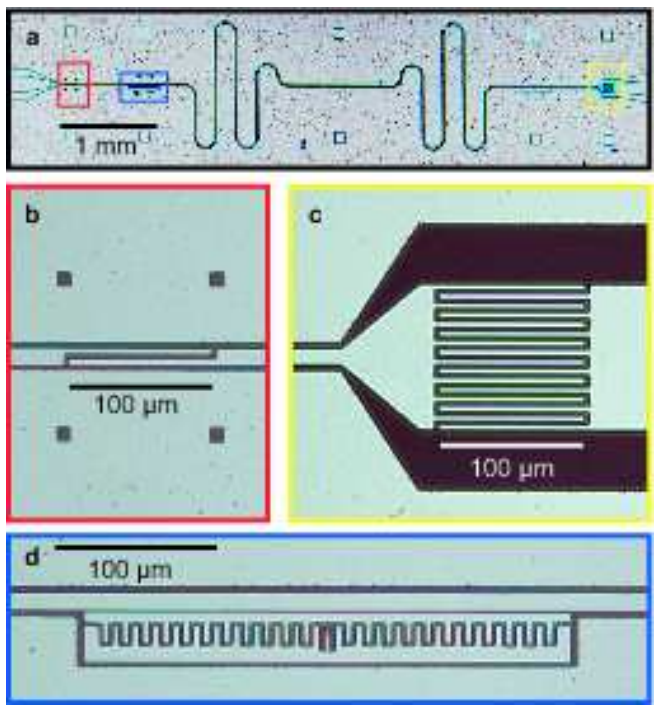}
\caption{The circuit QED device for generating single photons.
\textbf{a.} A transmission line cavity is formed between two
capacitors, with the input capacitor shown in \textbf{b} and the
output in \textbf{c}. Because the output is much larger, most
radiation leaving the cavity leaves from this port, allowing
efficient collection of light emitted from the cavity. \textbf{d.}
Transmon qubit, an optimized Cooper Pair Box, at a voltage anti-node
of the cavity. The qubit is characterized by a Josephson energy,
tuned by an applied magnetic field with a maximum of
$E_{\rm{J}}^{\rm{max}} = 20.2 \,\rm{GHz}$ and a charging energy
$E_{\rm{c}} = 0.37$ GHz. The coupling to the cavity is $g =
107\,\rm{MHz}$ at the qubit frequency primarily used in this paper,
$\omega_{\rm{a}} = 4.68\,\rm{GHz}$, and has a slight dependence on
the qubit frequency.}
\end{figure}

Verifying the single photon output is a substantial challenge in
on-chip microwave experiments. The simplest approach, looking for a
photon each time one is created, is not possible; unlike in optical
frequency experiments, no detectors can yet resolve single microwave
photon events in a single shot. Fortunately, several unique
characteristics of the source are evident in the average signal
generated by many single photon events, together yielding a
convincing verification even with noisy detectors.  Most simply, the
output of the single photon source is expected to be oscillatory in
the amplitude of the control pulse applied to rotate the qubit.
Second, the average amplitude produced should agree well with the
expected value for a single photon. Finally, and most importantly,
if the output of the system depends only on the state of the qubit,
state tomography measured for the photons should show complete
agreement with that obtained from independent measurements of the
qubit. The source reported here meets all three of these criteria.

The circuit designed to generate photons consists of a
superconducting transmon qubit\cite{koch07}, an optimized version of
the Cooper Pair Box\cite{bouchiat98}, capacitively coupled to a
half-wave transmission line cavity with fundamental frequency
$\omega_{\rm{r}}/2\pi = 5.19 \,\rm{GHz}$ (see Figure 1). Two
important design differences between this circuit and previous
incarnations of circuit QED\cite{wallraff04, schuster06} are needed
to achieve efficient single photon generation. First, the cavity is
asymmetric in that the capacitors (mirrors) at either end of the
transmission line are no longer equal, resulting in asymmetric decay
rates to the input and output ports ($\kappa_{\rm{in}}/2\pi \approx
200\,\rm{kHz}$ for the input side and $\kappa_{\rm{out}}/2\pi =
44\,\rm{MHz}$ for the output). As a result, photons generated in the
cavity are emitted at the output port more than $99\%$ of the time.
In addition, the total decay rate for the cavity, $\kappa/2\pi =
44\,\rm{MHz}$, is substantially higher than in previous samples, a
necessary change for spontaneous emission to be the dominant
relaxation channel for the qubit. The qubit decay rate in the
absence of spontaneous emission, $\gamma_{\rm{NR}}$, is frequency
dependent, with $\gamma_{\rm{NR}}/2\pi < 2\,\rm{MHz}$ for all
measured transmission frequencies between $4.3$ and $7.3\,\rm{GHz}$.

\begin{figure}[!bp]

\includegraphics[width = 0.48 \textwidth]{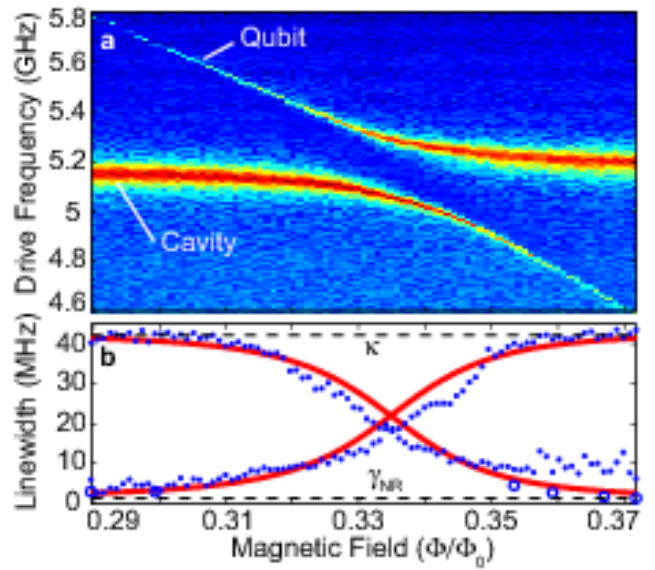}
\caption{Enhanced spontaneous emission through the Purcell effect.
\textbf{a.} Transmission through the cavity-qubit system at
different applied fluxes (log scale).  Two peaks are evident in
transmission due to the vacuum Rabi splitting.  Away from the
avoided crossing, these peaks correspond to ``mostly qubit'' and
``mostly cavity'' states. The bare linewidth of the cavity,
$\kappa/2\pi = 44\,\rm{MHz}$, is much larger than the bare qubit
linewidth $\gamma/2\pi < 2\,\rm{MHz}$. \textbf{b.} Extracted
linewidths from the data in (a) (closed circles) are compared with
theoretical values (red line). As the qubit and cavity peaks
approach degeneracy, the qubit peak becomes broader due to
spontaneous emission to the cavity mode, while the cavity decay is
suppressed. Extra dephasing present only at low frequencies (the
right side of the graph) causes a non-Lorentzian line shape and
excessive width. Measurements of the relaxation rate in the time
domain (open circles) agree with theoretical estimates.
Discrepancies arise due to flux instability and variations in
non-radiative decay with frequency.}
\end{figure}

Transmission measurements are used to probe the energy spectrum of
this system while the qubit frequency is tuned via an external
magnetic field (see Figure 2).  When the qubit is far detuned from
the cavity, only a single transmission peak is observed, centered at
the cavity frequency with a Lorentzian lineshape and width given by
the bare cavity width.  When the qubit and cavity are resonant, two
peaks in transmission are seen, a phenomenon known as the vacuum
Rabi splitting. Each peak corresponds to one of the two
single-excitation eigenstates of the system, superpositions of the
separate qubit and photon excitation states. The width of each peak
is the average of the qubit and photon decay rates, $(\gamma +
\kappa)/2$.  In the dispersive limit, where the detuning $\Delta$ is
much larger than the coupling $g$, spontaneous emission is enhanced
by the Purcell effect\cite{purcell46}, resulting in approximate
decay rates $[1-(g/\Delta)^2]\kappa + (g/\Delta)^2 \gamma$ and
$[1-(g/\Delta)^2]\gamma + (g/\Delta)^2 \kappa$. Experimentally
determined linewidths agree well with theoretical predictions
(Figure 2b), demonstrating an ability to tune the rate of radiative
decay of the qubit by tuning its frequency.

It is this enhanced qubit decay due to the cavity that is utilized
in generating single photons: the qubit line broadens when the decay
of the photon-like part of the wavefunction dominates the
non-radiative qubit decay.   For the results presented here, the
qubit was tuned to a frequency $\omega_{\rm{a}}/2\pi =
4.68\,\rm{GHz}$. With a coupling $g/2\pi = 107\,\rm{MHz}$, the qubit
wave functions had a $(g/\Delta)^2 = 4\%$ photonic nature, resulting
in a spontaneous emission rate $\gamma_{\kappa}/2\pi =
1.9\,\rm{MHz}$.  The measured relaxation rate of the qubit was
$\gamma/2\pi = 1.8 \pm 0.1\,\rm{MHz}$, indicating that the observed
relaxation could be mostly accounted for by spontaneous emission to
the cavity to within our measurement accuracy.  Because the lifetime
of the qubit is short, the photon source is effectively reset in
under $1\,\rm{\mu s}$, allowing for rapid repeated photon
generation, for a peak source power of $3\,\rm{aW}$.

\begin{figure}[!bp]

\includegraphics[width = 0.48 \textwidth]{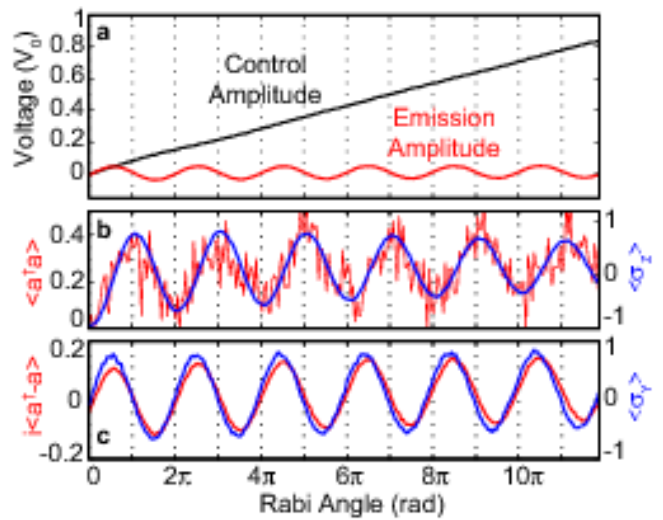}
\caption{Output of a single photon source.  \textbf{a.} Measured
drive and spontaneous emission voltage of cavity output, in units of
zero-point voltage fluctuations $V_0 \approx 2\,\rm{\mu V}$. When
the amplitude of the drive rotating the qubit is linearly increased,
the output voltage of the cavity is oscillatory. \textbf{b.} Output
power of the cavity $\left<a^{\dagger} a\right>$ detected with a
diode (left axis), and the measured qubit state
$\left<\sigma_{\rm{z}}\right>$ (right axis). These peak when the
qubit is in the excited state, after a $\pi$ pulse; the agreement
between qubit and photon states verifies the photon generation
occurs as expected. The power collected after a $36\,\rm{ns}$ delay
is $38\%$ of a single photon. Fits to the qubit decays at later
times are used to extrapolate the qubit polarization immediately
after the control pulse. \textbf{c.} Average voltage of the output
photons $i\left<a^{\dagger}-a\right>$ compared with the qubit state
$\left<\sigma_{\rm{y}}\right>$ measured with a Ramsey experiment.
The agreement shows that the phase of superposition states is also
transferred from qubit to photon. Only $12\%$ of the voltage for the
superposition is collected here, due non-radiative decay and
dephasing during the $36\,\rm{ns}$ delay after the control pulse.
The qubit amplitude is again extrapolated to the time immediately
following the control pulse.}
\end{figure}

To verify single photon generation, we first show that the output of
the cavity is an oscillatory function of the input drive, as at most
one photon is generated regardless of the magnitude of the input
drive. A $12\,\rm{ns}$ Gaussian control pulse rotates the qubit
state by a Rabi angle that is proportional to the pulse amplitude.
The excited qubit will relax, generating a new photon state at the
qubit frequency. Because the control pulse leaves the cavity at a
rate that is much faster than the rate of spontaneous emission
$\gamma_{\kappa}$, the control pulse and generated photons can
easily be separated in time.  As seen in Figure 3a, the measured
control signal increases monotonically, while the spontaneous
emission oscillates as the qubit is rotated through from the ground
to the excited state and back, confirming that the spontaneous
emission is proportional to the qubit state, not simply the applied
drive amplitude. This is the key to the experiment: a superposition
of many photons incoming on one temporal mode give rise to one and
only one photon on a distinct outgoing temporal mode.  Moreover,
because a single photon is the maximum output, the source is to
first order insensitive to fluctuations in the control pulse when
generating one photon.

We characterize both the power and electric field of the single
photon source, using independent measurements of the qubit state
made with dispersive readout technqiues\cite{wallraff05} to verify
performance (Figure 3). If the qubit state is mapped to the photon
state, then an arbitrary superposition of the ground and excited
states $\alpha \left|g\right> + \beta \left|e\right>$ will result in
the same superposition of photon states: $\alpha \left|0\right> +
\beta \left|1\right>$, where $\left|0\right>$ and $\left|1\right>$
refer to states with zero or one photon.  The average photon number
is proportional to the average qubit excitation probability,
$\left<a^{\dagger} a\right> = (\left<\sigma_{\rm{z}}\right>+1)/2$,
which has a maximum of one photon when the qubit is in the purely
excited state. The two quadratures of homodyne voltage, on the other
hand, are proportional to the x and y-components of the qubit state:
$\left<a + a^{\dagger}\right> = \left<\sigma_{\rm{x}}\right>$ and
$i\left<a^{\dagger}-a\right> = \left<\sigma_{\rm{y}}\right>$.  The
measured homodyne voltage is therefore a $\pi/2$ rotation
out-of-phase with the measured power, and the homodyne voltage is
zero when a single photon is generated, as there is complete phase
uncertainty in a photon Fock state.

Both the power and voltage of the photon output match the qubit
state, demonstrating the ability to generate single photons, as well
as arbitrary superpositions of zero and one photon, simply by
controlling the qubit.  Moreover, the amplitude of the pulse is as
expected.  The measured control pulse is used to calibrate the
amplitude of the spontaneous emission. The frequency of qubit
oscillations and a measurement of the control pulse on the output of
the cavity together yield a calibration for the gain of the
amplifiers, which in turn allows us to determine the efficiency of
our single photon source. Simulations that include non-radiative
channels of decay and dephasing agree well with the observed data
(see Figure 4), indicating that the power and amplitude of the
source are well understood.

\begin{figure}[!bp]

\includegraphics[width = 0.48 \textwidth]{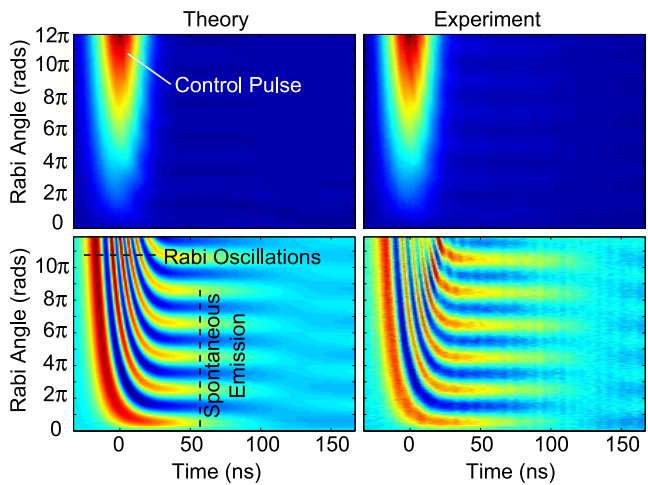}
\caption{Direct observation of the free induction decay of a
superconducting qubit. Theoretical predictions (left) for both
quadratures of the homodyne voltage, both in-phase (top) and
out-of-phase (bottom) with the drive, agree well with experimental
measurements of the two phases (right).  Because emission is always
orthogonal to the rotation axis, the spontaneous emission and
control signal are phase separable.  The homodyne sine waves in
Figure 3 represent an integral of vertical slices through the
emission. The frequency of these oscillations, coupled with a gain
known from measurements of the control pulse, provide a calibration,
which is used to predict the experimental emission data. Since the
qubit and drive are slightly detuned by a fluctuating amount due to
flux instability (on the order of $3\,\rm{MHz}$), there is a slow
beat note in the time direction.  This fluctuating detuning is
modeled by adding the predicted homodyne emission at two detunings,
$\pm 1.5\,\rm{MHz}$. The fast oscillations in the time domain are a
direct measure of the Rabi oscillations of the qubit. }
\end{figure}

Using this calibration technique and Markovian master equation
simulations, the complete time dynamics can be predicted to
excellent accuracy, as shown for the homodyne voltage in Figure 4.
Several features of the time dynamics are striking.  First, because
the control pulse sets the rotation axis, and the qubit state sets
the emission phase, the control and generated photons are orthogonal
in phase, which allows the two signals to be completely separated in
homodyne detection. In the generated photon quadrature, rapid time
oscillations are apparent during the control pulse; this is a direct
observation of the Rabi oscillation of the qubit through its
spontaneous emission. After the pulse, the qubit emits with a phase
depending on its final state, resulting in oscillations in the
control amplitude that smoothly connect to the time oscillations.
Finally, there is a very low frequency oscillation in time. Photons
are emitted at the qubit frequency, which is slightly detuned from
the drive frequency. The result is a beating, with a half period
shown in both theory and data images, indicative of a frequency
separation between the input and output photons in addition to the
phase and time separations.

\begin{figure}[!bp]
\includegraphics[width = 0.48\textwidth]{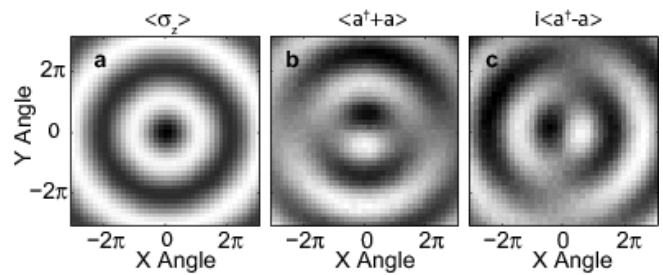}
\caption{Mapping the qubit state onto the photon state. \textbf{a.}
Measurement of qubit state $\left<\sigma_{\rm{z}}\right>$ after
rotations by pulses of arbitrary amplitude and phase. Regardless of
the phase of the pulse, the qubit oscillates to a peak after a $\pi$
pulse. \textbf{b} and \textbf{c.} Fluorescence tomography.  The
amplitude of the voltage measured in each homodyne quadrature,
$\left<a + a^{\dagger}\right>$ and $i\left<a^{\dagger}-a\right>$,
agree with expectations for $\left<\sigma_{\rm{x}}\right>$ and
$\left<\sigma_{\rm{y}}\right>$. Oscillations around the x-axis
produce a signal in $i\left<a^{\dagger}-a\right>$ and none in
$\left<a + a^{\dagger}\right>$.  This shows the ability to map an
arbitrary qubit state onto a photon state, as well as the ability to
characterize a qubit state through spontaneous emission. }
\end{figure}

Two metrics of efficiency characterize the performance of the
system.  The source efficiency is the fraction of time in which a
photon is emitted after a control pulse.  This depends on the final
polarization of the qubit and the ratio of radiative to
non-radiative channels.  In generating a single photon, the
$\pi$-pulse leaves the qubit $87\%$ polarized, and nearly all decay
is radiative, giving source efficiency close to $87\%$.  For
generating a superposition of zero and one photon, the quadrature
phase of the photon must also be controlled.  Here, the qubit is
$77\%$ polarized along $\sigma_{\rm{y}}$, but a dephasing rate
$\gamma_{\phi} = 1\,\rm{MHz}$ leads to only
$\gamma_{\kappa}/(2\gamma_{\phi}+\gamma_{\kappa}) = 50\%$ radiation
efficiency, giving a total source efficiency of $39\%$.

A second metric, the \emph{usable} source efficiency, is somewhat
lower in the current experiment, as the control pulse is slow and a
delay is necessary to reject any control photons which could give a
double-photon event.  In the data of Figure 3, collection of
radiation begins after $3$ standard deviations of the control pulse
Gaussian, making the likelihood of a control photon less than
$0.01\%$. Integrating the measured spontaneous emission, the number
of detected photons is measured directly, yielding an efficiency of
$38\%$ for the single photon source, and $12\%$ for the homodyne
voltage of the superposition state, which is again lower due to
dephasing. Even rejecting the emission contaminated by control pulse
photons, which contains the high signal-to-noise part of the
emission, a substantial fraction of one photon is recovered.  If a
less stringent rejection threshold of $1\%$ is chosen, efficiencies
rise to $46\%$ for power and $16\%$ for homodyne voltage.  In future
experiments, this could be improved further with faster pulses,
longer coherence times, or fast tunability of the qubit frequency,
achieving usable source efficiencies close to $100\%$.

Tomography presents an even more powerful tool for characterizing
the qubit\cite{steffen06} and photon states\cite{smithey93,
leonhardt97}, and demonstrates the complete mapping of the qubit
state onto the photons (Figure 5). Here, qubit tomography is
performed by applying control pulses of arbitrary phase and
amplitude, and performing a dispersive measurement of the qubit
state $\left<\sigma_{\rm{z}}\right>$. This yields the expected
concentric rings for a qubit initially in the ground state (Figure
5a). A fluorescence homodyne tomography technique is used to
characterize the photons.  Control pulses with all phases and
amplitudes are applied, and both quadratures of the output homodyne
voltage are recorded (Figures 5b and 5c). These show excellent
agreement with the expected $\sigma_{\rm{x}}$ and $\sigma_{\rm{y}}$
components of the qubit state\cite{steffen06}. This fluorescence
tomography technique allows a full characterization of the qubit by
looking at the spontaneous emission at the output, directly
observing a superconducting qubit at its Larmor frequency. Moreover,
this means that a qubit state can be transferred onto a photon
state, thus moving information from a stationary qubit to a ``flying
qubit'', one of the DiVincenzo resources for quantum information
processing\cite{divincenzo00}.

The mapping of qubit states onto photon states allows for the use of
microwave photons as a true resource for quantum information on a
chip. These photons are generated on-demand with a high repetition
rate, good spectral purity, and high efficiency.  This is a
convenient means of creating non-classical states of light to
interact with atoms, all in the wires of an integrated circuit,
allowing them to be shuttled around a chip.  The generation of
single photons, and superpositions of photon states, is an important
step towards on-chip quantum optics experiments.


\begin{thebibliography}{10}

\bibitem{schuster07}
\bibinfo{author}{Schuster, D.~I.} \emph{et~al.}
\newblock \bibinfo{title}{Resolving photon number states in a superconducting
  circuit}.
\newblock \emph{\bibinfo{journal}{Nature}} \textbf{\bibinfo{volume}{445}},
  \bibinfo{pages}{515--518} (\bibinfo{year}{2007}).


\bibitem{divincenzo00}
\bibinfo{author}{DiVincenzo, D.~P.}
\newblock \bibinfo{title}{The physical implementation of quantum computation}.
\newblock \emph{\bibinfo{journal}{Fortschritte der Physik}}
  \textbf{\bibinfo{volume}{48}}, \bibinfo{pages}{771--783}
  (\bibinfo{year}{2000}).

\bibitem{blais04}
\bibinfo{author}{Blais, A.}, \bibinfo{author}{Huang, R.-S.},
  \bibinfo{author}{Wallraff, A.}, \bibinfo{author}{Girvin, S.~M.} \&
  \bibinfo{author}{Schoelkopf, R.~J.}
\newblock \bibinfo{title}{Cavity quantum electrodynamics for superconducting
  electrical circuits: An architecture for quantum computation}.
\newblock \emph{\bibinfo{journal}{Phys. Rev. A}} \textbf{\bibinfo{volume}{69}},
  \bibinfo{pages}{062320--14} (\bibinfo{year}{2004}).

\bibitem{oxborrow05}
\bibinfo{author}{Oxborrow, M.} \& \bibinfo{author}{Sinclair, A.}
\newblock \bibinfo{title}{Single-photon sources}.
\newblock \emph{\bibinfo{journal}{Contemporary Physics}}
  \textbf{\bibinfo{volume}{46}}, \bibinfo{pages}{173--206}
  (\bibinfo{year}{2005}).

\bibitem{clauser74}
\bibinfo{author}{Clauser, J.~F.}
\newblock \bibinfo{title}{Experimental distinction between the quantum and
  classical field-theoretic predictions for the photoelectric effect}.
\newblock \emph{\bibinfo{journal}{Phys. Rev. D}} \textbf{\bibinfo{volume}{9}},
  \bibinfo{pages}{853--} (\bibinfo{year}{1974}).

\bibitem{kimble77}
\bibinfo{author}{Kimble, H.~J.}, \bibinfo{author}{Dagenais, M.} \&
  \bibinfo{author}{Mandel, L.}
\newblock \bibinfo{title}{Photon antibunching in resonance fluorescence}.
\newblock \emph{\bibinfo{journal}{Phys. Rev. Lett.}}
  \textbf{\bibinfo{volume}{39}}, \bibinfo{pages}{691--} (\bibinfo{year}{1977}).

\bibitem{diedrich87}
\bibinfo{author}{Diedrich, F.} \& \bibinfo{author}{Walther, H.}
\newblock \bibinfo{title}{Nonclassical radiation of a single stored ion}.
\newblock \emph{\bibinfo{journal}{Phys. Rev. Lett.}}
  \textbf{\bibinfo{volume}{58}}, \bibinfo{pages}{203--} (\bibinfo{year}{1987}).

\bibitem{darquie05}
\bibinfo{author}{Darquie, B.} \emph{et~al.}
\newblock \bibinfo{title}{Controlled single-photon emission from a single
  trapped two-level atom}.
\newblock \emph{\bibinfo{journal}{Science}} \textbf{\bibinfo{volume}{309}},
  \bibinfo{pages}{454--456} (\bibinfo{year}{2005}).

\bibitem{basche92}
\bibinfo{author}{Basche, T.}, \bibinfo{author}{Moerner, W.~E.},
  \bibinfo{author}{Orrit, M.} \& \bibinfo{author}{Talon, H.}
\newblock \bibinfo{title}{Photon antibunching in the fluorescence of a single
  dye molecule trapped in a solid}.
\newblock \emph{\bibinfo{journal}{Phys. Rev. Lett.}}
  \textbf{\bibinfo{volume}{69}}, \bibinfo{pages}{1516--}
  (\bibinfo{year}{1992}).

\bibitem{brunel99}
\bibinfo{author}{Brunel, C.}, \bibinfo{author}{Lounis, B.},
  \bibinfo{author}{Tamarat, P.} \& \bibinfo{author}{Orrit, M.}
\newblock \bibinfo{title}{Triggered source of single photons based on
  controlled single molecule fluorescence}.
\newblock \emph{\bibinfo{journal}{Phys. Rev. Lett.}}
  \textbf{\bibinfo{volume}{83}}, \bibinfo{pages}{2722--}
  (\bibinfo{year}{1999}).

\bibitem{lounis00}
\bibinfo{author}{Lounis, B.} \& \bibinfo{author}{Moerner, W.~E.}
\newblock \bibinfo{title}{Single photons on demand from a single molecule at
  room temperature}.
\newblock \emph{\bibinfo{journal}{Nature}} \textbf{\bibinfo{volume}{407}},
  \bibinfo{pages}{491--493} (\bibinfo{year}{2000}).

\bibitem{kurtsiefer00}
\bibinfo{author}{Kurtsiefer, C.}, \bibinfo{author}{Mayer, S.},
  \bibinfo{author}{Zarda, P.} \& \bibinfo{author}{Weinfurter, H.}
\newblock \bibinfo{title}{Stable solid-state source of single photons}.
\newblock \emph{\bibinfo{journal}{Phys. Rev. Lett.}}
  \textbf{\bibinfo{volume}{85}}, \bibinfo{pages}{290--} (\bibinfo{year}{2000}).

\bibitem{michler00_nature}
\bibinfo{author}{Michler, P.} \emph{et~al.}
\newblock \bibinfo{title}{Quantum correlation among photons from a single
  quantum dot at room temperature}.
\newblock \emph{\bibinfo{journal}{Nature}} \textbf{\bibinfo{volume}{406}},
  \bibinfo{pages}{968--970} (\bibinfo{year}{2000}).

\bibitem{pelton02}
\bibinfo{author}{Pelton, M.} \emph{et~al.}
\newblock \bibinfo{title}{Efficient source of single photons: A single quantum
  dot in a micropost microcavity}.
\newblock \emph{\bibinfo{journal}{Phys. Rev. Lett.}}
  \textbf{\bibinfo{volume}{89}}, \bibinfo{pages}{233602--}
  (\bibinfo{year}{2002}).

\bibitem{purcell46}
\bibinfo{author}{Purcell, E.~M.}
\newblock \bibinfo{title}{Spontaneous emission probabilities at radio
  frequencies}.
\newblock \emph{\bibinfo{journal}{Physical Review}}
  \textbf{\bibinfo{volume}{69}}, \bibinfo{pages}{681} (\bibinfo{year}{1946}).

\bibitem{brattke01}
\bibinfo{author}{Brattke, S.}, \bibinfo{author}{Varcoe, B. T.~H.} \&
  \bibinfo{author}{Walther, H.}
\newblock \bibinfo{title}{Generation of photon number states on demand via
  cavity quantum electrodynamics}.
\newblock \emph{\bibinfo{journal}{Phys. Rev. Lett.}}
  \textbf{\bibinfo{volume}{86}}, \bibinfo{pages}{3534--}
  (\bibinfo{year}{2001}).

\bibitem{kuhn02}
\bibinfo{author}{Kuhn, A.}, \bibinfo{author}{Hennrich, M.} \&
  \bibinfo{author}{Rempe, G.}
\newblock \bibinfo{title}{Deterministic single-photon source for distributed
  quantum networking}.
\newblock \emph{\bibinfo{journal}{Phys. Rev. Lett.}}
  \textbf{\bibinfo{volume}{89}}, \bibinfo{pages}{067901--}
  (\bibinfo{year}{2002}).

\bibitem{mckeever04}
\bibinfo{author}{McKeever, J.} \emph{et~al.}
\newblock \bibinfo{title}{Deterministic generation of single photons from one
  atom trapped in a cavity}.
\newblock \emph{\bibinfo{journal}{Science}} \textbf{\bibinfo{volume}{303}},
  \bibinfo{pages}{1992--1994} (\bibinfo{year}{2004}).

\bibitem{maitre97}
\bibinfo{author}{Maitre, X.} \emph{et~al.}
\newblock \bibinfo{title}{Quantum memory with a single photon in a cavity}.
\newblock \emph{\bibinfo{journal}{Phys. Rev. Lett.}}
  \textbf{\bibinfo{volume}{79}}, \bibinfo{pages}{769--} (\bibinfo{year}{1997}).

\bibitem{keller04}
\bibinfo{author}{Keller, M.}, \bibinfo{author}{Lange, B.},
  \bibinfo{author}{Hayasaka, K.}, \bibinfo{author}{Lange, W.} \&
  \bibinfo{author}{Walther, H.}
\newblock \bibinfo{title}{Continuous generation of single photons with
  controlled waveform in an ion-trap cavity system}.
\newblock \emph{\bibinfo{journal}{Nature}} \textbf{\bibinfo{volume}{431}},
  \bibinfo{pages}{1075--1078} (\bibinfo{year}{2004}).

\bibitem{moreau01}
\bibinfo{author}{Moreau, E.} \emph{et~al.}
\newblock \bibinfo{title}{Single-mode solid-state single photon source based on
  isolated quantum dots in pillar microcavities}.
\newblock \emph{\bibinfo{journal}{Appl. Phys. Lett.}}
  \textbf{\bibinfo{volume}{79}}, \bibinfo{pages}{2865--2867}
  (\bibinfo{year}{2001}).

\bibitem{santori02}
\bibinfo{author}{Santori, C.}, \bibinfo{author}{Fattal, D.},
  \bibinfo{author}{Vuckovic, J.}, \bibinfo{author}{Solomon, G.~S.} \&
  \bibinfo{author}{Yamamoto, Y.}
\newblock \bibinfo{title}{Indistinguishable photons from a single-photon
  device}.
\newblock \emph{\bibinfo{journal}{Nature}} \textbf{\bibinfo{volume}{419}},
  \bibinfo{pages}{594--597} (\bibinfo{year}{2002}).

\bibitem{wallraff04}
\bibinfo{author}{Wallraff, A.} \emph{et~al.}
\newblock \bibinfo{title}{Strong coupling of a single photon to a
  superconducting qubit using circuit quantum electrodynamics}.
\newblock \emph{\bibinfo{journal}{Nature}} \textbf{\bibinfo{volume}{431}},
  \bibinfo{pages}{162--167} (\bibinfo{year}{2004}).

\bibitem{chiorescu04}
\bibinfo{author}{Chiorescu, I.} \emph{et~al.}
\newblock \bibinfo{title}{Coherent dynamics of a flux qubit coupled to a
  harmonic oscillator}.
\newblock \emph{\bibinfo{journal}{Nature}} \textbf{\bibinfo{volume}{431}},
  \bibinfo{pages}{159--162} (\bibinfo{year}{2004}).

\bibitem{koch07}
\bibinfo{author}{Koch, J.} \emph{et~al.}
\newblock \bibinfo{title}{Optimizing the cooper pair box: Introducing the
  transmon}.
\newblock \bibinfo{howpublished}{In preparation} (\bibinfo{year}{2007}).

\bibitem{bouchiat98}
\bibinfo{author}{Bouchiat, V.}, \bibinfo{author}{Vion, D.},
  \bibinfo{author}{Joyez, P.}, \bibinfo{author}{Esteve, D.} \&
  \bibinfo{author}{Devoret, M.~H.}
\newblock \bibinfo{title}{Quantum coherence with a single cooper pair}.
\newblock \emph{\bibinfo{journal}{Physica Scripta}}
  \textbf{\bibinfo{volume}{T76}}, \bibinfo{pages}{165--170}
  (\bibinfo{year}{1998}).

\bibitem{wallraff05}
\bibinfo{author}{Wallraff, A.} \emph{et~al.}
\newblock \bibinfo{title}{Approaching unit visibility for control of a
  superconducting qubit with dispersive readout}.
\newblock \emph{\bibinfo{journal}{Phys. Rev. Lett.}}
  \textbf{\bibinfo{volume}{95}}, \bibinfo{pages}{060501--4}
  (\bibinfo{year}{2005}).

\bibitem{steffen06}
\bibinfo{author}{Steffen, M.} \emph{et~al.}
\newblock \bibinfo{title}{State tomography of capacitively shunted phase qubits
  with high fidelity}.
\newblock \emph{\bibinfo{journal}{Phys. Rev. Lett.}}
  \textbf{\bibinfo{volume}{97}}, \bibinfo{pages}{050502--4}
  (\bibinfo{year}{2006}).

\bibitem{smithey93}
\bibinfo{author}{Smithey, D.~T.}, \bibinfo{author}{Beck, M.},
  \bibinfo{author}{Raymer, M.~G.} \& \bibinfo{author}{Faridani, A.}
\newblock \bibinfo{title}{Measurement of the wigner distribution and the
  density matrix of a light mode using optical homodyne tomography: Application
  to squeezed states and the vacuum}.
\newblock \emph{\bibinfo{journal}{Phys. Rev. Lett.}}
  \textbf{\bibinfo{volume}{70}}, \bibinfo{pages}{1244--}
  (\bibinfo{year}{1993}).

\bibitem{leonhardt97}
\bibinfo{author}{Leonhardt, U.}
\newblock \emph{\bibinfo{title}{Measuring the Quantum State of Light}}
  (\bibinfo{publisher}{Cambridge University Press}, \bibinfo{year}{1997}).

\end{thebibliography}



\begin{acknowledgments}
 This work was supported in part by the National Security Agency
   under the Army Research Office, the NSF, and Yale University.
   A.H. would like to acknowledge support from Yale
   University via a Quantum Information and Mesoscopic Physics
   Fellowship.
The authors declare that they have no competing financial interests.
Correspondence and requests for materials should be addressed to Rob
Schoelkopf~(email:Robert.Schoelkopf@yale.edu).
\end{acknowledgments}

\end{document}